\documentstyle[epsfig]{lamuphys}
\makeatletter
\let\chapter\hid@chapter
\makeatother
\begin{document}
\pagenumbering{arabic}
\title{Constraining the Mass Distribution of Cluster Galaxies by Weak Lensing}

\author{Bernhard Geiger}

\institute{Max-Planck-Institut f\"ur Astrophysik, 
Karl-Schwarzschild-Stra{\ss}e 1, Postfach 1523, 
D-85740 Garching bei M\"unchen, Germany}

\maketitle

\begin{abstract}
Analysing the weak lensing distortions of the images of faint background
galaxies provides a means to constrain the mass distribution of cluster
galaxies and potentially to test the extent of their dark matter halos as a
function of the density of the environment. Here I describe simulations of
observational data and present a maximum likelihood method to infer the average
properties of an ensemble of cluster galaxies.   

\end{abstract}
\section{Introduction}
Measurements of the rotation curves of spiral galaxies indicate that they are
embedded in massive dark matter halos. The deflection of light rays through the
gravitational action of mass concentrations, usually called gravitational
lensing, provides a way to obtain information about the mass distribution of
galaxies at radial distances from their centre where there are no more luminous
test particles to probe the gravitational potential. The light deflection
causes small distortions of the images of faint background galaxies. Recent
statistical analyses (Brainerd et al. 1996, Griffiths et al. 1996) of these
weak distortion effects suggest that the dark galaxy halos are indeed rather
extended, as some popular theories of structure formation predict them to
be. During the formation of galaxy clusters the extended halos of galaxies may
be stripped off due to tidal forces of the cluster potential or during
encounters with other galaxies. Ultimately the individual galaxy halos should
merge and form a global cluster halo. In this contribution I discuss how this
merging picture could be tested observationally by exploiting the weak lensing
effects.   

The distortions of the images of background galaxies produced by massive galaxy
clusters are strong enough to allow a parameter-free reconstruction of
the clusters' surface mass density, and several algorithms have been developed
for this purpose (e.g. Kaiser and Squires 1993, Seitz and Schneider 1995,
1996). The smoothing length which has to be implemented in these techniques,
however, is larger than galaxy scales, i.e., the amount of information
available does not suffice to reconstruct cluster galaxies
individually. Therefore, one has to superpose the effects of a large number of 
galaxies statistically in order to infer the average properties of an ensemble
of galaxies.  

Section\,\ref{simulations} presents simulations of a galaxy cluster which are
sufficiently realistic for the purposes of this work, and demonstrates how
individual galaxies modify the distortion pattern of a smooth cluster mass
distribution. Section\,\ref{method} discusses a maximum likelihood
method for constraining the mass distribution of cluster galaxies, and
Sect.\,\ref{results} presents results of the simulations. Finally, in
Sect.\,\ref{prospects} some suggestions for refining the simulations are
mentioned, and observational prospects are discussed. A closely related work
was recently published by Natarajan and Kneib (1997); in contrast to their
maximum likelihood method, the mass profile of the cluster is 
not assumed to be known but is reconstructed from image distortions as
mentioned above.  

\section{Simulations}\label{simulations}
\subsection{Cluster and Cluster Galaxies}
\begin{figure}
  \begin{center}
    \epsfysize=5cm
    \epsffile{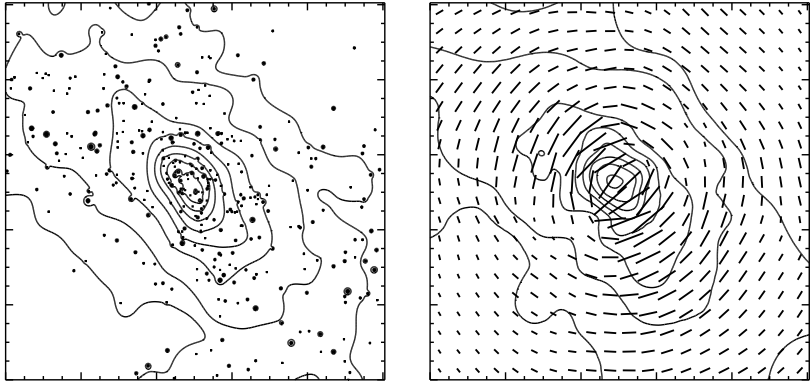}
  \end{center}
\caption[]{\small Left: The mass distribution of the cluster including
the cluster galaxies with $s_{\star}=34h^{-1}\,\rm{kpc}$. Right: The
distortion pattern determined from the ellipticities of background galaxy
images overlaid with the reconstructed cluster mass distribution. The field of
view is $10\arcmin$ and the contours are
$\kappa=0.05,0.1,0.2,0.3,0.4,0.5,0.6,\mbox{ and }0.7$.}  
\label{cluster}
\end{figure}

A galaxy cluster with a total mass of about $10^{15}h^{-1}\,\rm{M}_{\sun}$
located at a redshift of $z_d=0.16$ was selected from numerical N-body
simulations (Bartelmann et al. 1995). Within this paper a quadratic field of
view with side length $10\arcmin$ is considered, which roughly corresponds to a
physical size of $1h^{-1}\,\rm{Mpc}$ at the cluster redshift. In order to
populate the dark matter distribution of this cluster with galaxies the
following requirements were specified:
\begin{enumerate}
\item The total mass-to-light ratio of the cluster was chosen to be
$300h\,\rm{M}_{\sun}/\rm{L}_{\sun}$.
\item Galaxy luminosities $L$ were drawn from a Schechter function with
canonical parameters (and a cutoff at $0.1\,L_{\star}$).
\item Galaxy positions were randomly drawn from those of the N-body particles.
\end{enumerate}
This procedure resulted in a rich cluster of 359 galaxies, 40 of which are
brighter than $L_{\star}$. For the mass distribution of the cluster
galaxies, a simple truncated isothermal sphere (Brainerd et al. 1996) was used.
The surface mass density $\Sigma$ as a function of the projected radius $\xi$
is given by 
\begin{equation}\label{iso}
\Sigma(\xi)=\frac{\sigma^2}{2\rm{G}\xi}\,
\left(1 - \frac{\xi}{\sqrt{s^2+\xi^2}}\right)\,,
\end{equation}
where the two parameters, velocity dispersion $\sigma$ and cutoff radius $s$,
were chosen as functions of the luminosity according to the following scaling
relations:
\begin{equation}\label{scal}
\sigma=\sigma_{\star}\,\left(\frac{L}{L_{\star}}\right)^{1/\eta}\,\,\,
\mbox{and}\,\,\,\,\,\,s=s_{\star}\,\left(\frac{L}{L_{\star}}\right)^{\nu}.
\end{equation}
For the first of these relations, which is motivated by the observed
Tully-Fisher and Faber-Jackson relations, a value of $\eta=4$ was used for the 
scaling index and the velocity dispersion $\sigma_{\star}$ of an
$L_{\star}$-galaxy was fixed at $200\,\rm{km/s}$. For sim\-plicity, no
distinction between spiral and elliptical galaxies was made. The scaling
relation for the cutoff radius is more conjectural, and choosing $\nu=0.5$
yields a mass-to-light ratio for the galaxies which is independent of
luminosity. To test the method, two models were used for the cutoff
radius. Choosing $s_{\star}=3.4h^{-1}\,\rm{kpc}$ gives a total
$L_{\star}$-galaxy mass of $M_{\star}=10^{11}h^{-1}\,\rm{M}_{\sun}$, whereas an
extended halo of $s_{\star}=34h^{-1}\,\rm{kpc}$ results in
$M_{\star}=10^{12}h^{-1}\,\rm{M}_{\sun}$. These galaxy mass models were
added to the global cluster mass distribution, which was scaled such that the
total mass of the system remains constant (see Fig.\,\ref{cluster} left). 

\subsection{Distortion Effects and Background Galaxies}
The lensing properties of the galaxy cluster are specified by the dimensionless
surface mass density $\kappa$ and the (complex) shear $\gamma$ which are second
derivatives of a common two-dimensional scalar potential. However, image
distortions are only sensitive to the combined quantity $g=\gamma/(1-\kappa)$. 
Figure\,\ref{distortion} shows a map of $|g|$, which is a measure of the
strength of the distortion effects on the images of background galaxies. In
general, these distortions tend to be aligned tangentially towards the centre
of mass concentrations. The figure illustrates the perturbing effects of the
individual cluster galaxies. At their positions in a radial direction towards
and away from the cluster centre the strength of the distortion is locally
increased because the effects of the global cluster mass distribution and the
cluster galaxy then act in the same direction. But in the direction tangential
to the cluster centre the orientation of the galaxy contribution to the shear
is perpendicular to the cluster's shear direction, and therefore these effects
cancel out which leads to a reduction in the strength of the distortion
effects.   

\begin{figure}
\vspace{13cm}
\caption[]{\small The modulus $|g|$ of the reduced shear. This quantity is a
measure for the strength of the distortion effects.}
\label{distortion}
\end{figure}

Unfortunately Nature does not provide us with a continuous map of the lensing
properties, but only with very noisy estimates of the parameter $g$ at the
discrete positions of background galaxy images. For these simulations, a
random population of background galaxies was generated with a number density of
$40/\rm{arcmin}^2$, including a realistic intrinsic ellipticity distribution
and a reasonable redshift distribution. Figure\,\ref{cluster} (right) shows
the gridded distortion pattern calculated from the `observed' ellipticities of
the background galaxy images by employing a suitable averaging procedure. This
figure also displays the reconstruction of the mass distribution using a
finite-field non-linear inversion method (Seitz and Schneider 1996).

\section{Method}\label{method}
In order to constrain the mass distribution of cluster galaxies a maximum
likelihood method was developed which follows in part the prescription
of Schneider and Rix (1997) for weak lensing by field galaxies. The image
distortions are a consequence of the interplay between the effects of a global
cluster potential and the perturbations due to individual galaxies. In addition
to specifying a parametrized mass model for the galaxies it is, therefore,
important to have an accurate description of the cluster mass distribution
which is provided by the reconstruction mentioned above. As a model for the
galaxy mass distribution I again used the truncated isothermal sphere
(\ref{iso}). Of course, this model is appropriate for the simulated data used
here, whereas one could argue that realistic galaxy halos in clusters might
rather be flattened or completely irregular. However, this analysis is aimed at
determining the average properties of an ensemble of galaxies which might still
be reasonably described by a simple model with a characteristic scale and
normalization as parameters. In order to add the information from galaxies with
different luminosities, the scaling relations (\ref{scal}) were applied. Adding
the mass models for each of the cluster galaxies to the cluster reconstruction
then yields a model for the total mass distribution of the system as a function
of the velocity dispersion $\sigma_{\star}$ and the cutoff radius $s_{\star}$
of an $L_{\star}$-galaxy. A complication which has to be taken into account
when performing this procedure is the following. If the individual galaxies do
have extended halos, the mass in galaxies constitutes a significant fraction of
the total cluster mass ($\approx30\%$ for the model with 
$s_{\star}=34h^{-1}\,\rm{kpc}$). The cluster reconstruction is sensitive to the
total mass and therefore it already includes the masses of the galaxies. This
means that the additional mass added by the galaxy models has to be compensated
in some way. This was done by simply scaling down the reconstruction
appropriately or by subtracting surplus mass locally on scales of roughly
$1\arcmin$ at the position of cluster galaxies. The merits and limitations of
this ({\it ad hoc}\/) procedure will be discussed in Sect.\,\ref{results}.   

The total mass model constructed in this way determines the values for the
lensing parameters $\kappa$ and $\gamma$ at the position of each background
galaxy image. The strength of the lensing effect also depends on the distance
of the sources, and in the following the symbols $\kappa_{\infty}$ and
$\gamma_{\infty}$ are used to indicate the reference to (hypothetical) sources
located at infinite redshift. The probability density
$p_{\epsilon}(\epsilon\,|\,\kappa_{\infty},\,\gamma_{\infty})$ for observing 
the (complex) image ellipticity $\epsilon$ if the source is lensed by the
specified mass model is given by 
\begin{equation}
p_{\epsilon}(\epsilon\,|\,\kappa_{\infty},\,\gamma_{\infty})
=\int\limits_0^{_{\infty}}dz\,p_z(z)\,
p_{\epsilon_s}(\epsilon_s(\epsilon\,|\,\kappa_{\infty},\,\gamma_{\infty},\,z))
\,\left|\frac{\D^2\epsilon_s}{\D^2\epsilon}\right|
(\epsilon\,|\,\kappa_{\infty},\,\gamma_{\infty},\,z)\,.
\end{equation}
To calculate this probability, it is necessary to know the intrinsic
ellipticity distribution $p_{\epsilon_s}(\epsilon_s)$ of the sources, which can
be determined from `empty' fields, and an estimate for the redshift
distribution $p_z(z)$. In addition, non-linear properties of the lens mapping
have to be taken into account, and the last term under the integral is the
Jacobian determinant for the transformation of image ellipticities $\epsilon$
to source ellipticities $\epsilon_s$. The likelihood function ${\cal L}$ is
defined as the product of the probability densities of the actually
measured ellipticities $\epsilon_i$ of all the background galaxy images:
\begin{equation}
{\cal L}=\prod_i\,p(\epsilon_i)\,.
\end{equation}
The best-fit galaxy mass distribution and confidence regions can then be found
by maximizing this likelihood function with respect to the parameters of the
model. The logarithm of the likelihood function is denoted as 
$l=\ln{\cal L}$, and in the next section contour plots of 
$\Delta l=l-l_{\rm{Max}}$ as a function of the model parameters will be
presented.   

\section{Results}\label{results}
\subsection{Velocity Dispersion and Cutoff Radius}
\begin{figure}
  \begin{center}
    \epsfysize=4.5cm
    \epsffile{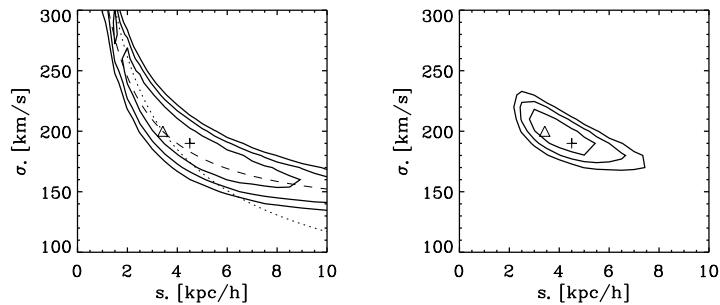}
  \end{center}
\caption[]{\small The logarithm of the likelihood as a function of the velocity
dispersion $\sigma_{\star}$ and the cutoff radius $s_{\star}$. The contours are
$\Delta l=-1,-2,-3$. The triangles denote the input values and the crosses mark
the maximum of the likelihood function. {\bf Left:} Only including information
provided by the lensing analysis. The dotted line connects models with equal
total mass and along the dashed line the mass within a projected radius of
$6h^{-1}\,\rm{kpc}$ is constant. {\bf Right:} The likelihood contours after
adding the prior information of $\sigma_{\star}=200\pm20\,\rm{km/s}$.} 
\label{like}
\end{figure}

Figure\,\ref{like} (left) shows the result of the maximum likelihood
analysis for the input galaxy model with a small cutoff radius of
$s_{\star}=3.4h^{-1}\,\rm{kpc}$. For this plot a total number of 3978
background galaxies from the entire field of view were included in the
calculation of the likelihood function. The confidence region closely follows
the dashed line of models with equal mass within a (projected) radius of 
$6h^{-1}\,\rm{kpc}$, which means that this is the quantity which can be
determined best with this lensing method. However, the velocity dispersion
and the cutoff radius individually cannot be well determined from the data. 
This result is not surprising because all background galaxy images which are
located closer to cluster galaxies than roughly this distance of
$6h^{-1}\,\rm{kpc}$ have been excluded from the analysis as they will be
outshone by the light of the cluster galaxy, and so there is no information
available about the distribution of the mass inside this radius.
But still it is possible to obtain tighter limits on the cutoff radius by
including {\it a priori}\/ knowledge on the velocity dispersion
$\sigma_{\star}$. If we believe that the measured velocity dispersions of
elliptical galaxies or the rotational velocities of spirals (divided by a
factor of $\sqrt{2}$) represent the same quantity as the parameter $\sigma$ of
the dark matter halo model, we can include this knowledge into the likelihood
function. Figure\,\ref{like} (right) demonstrates the results after adding
the prior information of $\sigma_{\star}=200\pm20\,\rm{km/s}$. In this case
very interesting limits on $s_{\star}$ can be achieved. 

\begin{figure}
  \begin{center}
    \epsfysize=4.5cm
    \epsffile{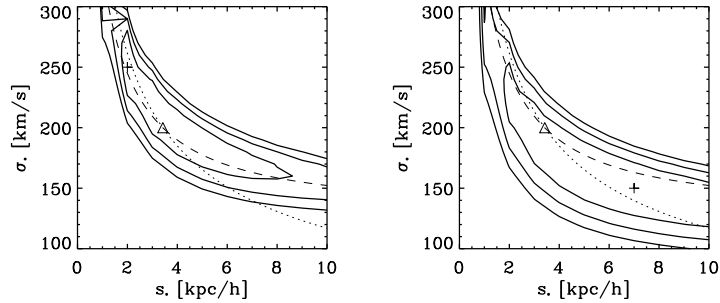}
  \end{center}
\caption[]{The same as Fig.\,\ref{like} (left) but after dividing the data set
into background galaxies with a distance from the cluster centre larger than
$1\arcmin\!.5$ ({\bf left}) and less than $1\arcmin\!.5$ ({\bf right}).} 
\label{small}
\end{figure}

In Fig.\,\ref{small} the same data is divided into two independent subsets
according to the location of the background galaxy images. The left panel shows
the contours of the logarithm of the likelihood function (without prior
information) using all images (3714) whose distances from the cluster centre
exceed $1\arcmin\!.5$ and in the right panel all images (264) within this limit
were used. The number of cluster galaxies which are located in these areas are
265 and 94, respectively. The figure shows that the far fewer images in the
centre provide almost the same amount of information as the numerous images in
the outskirts of the cluster. The reasons for this are the higher cluster
galaxy density in the centre and the significant enhancement of the distortion
effects of individual cluster galaxies due to the underlying cluster mass
distribution. Hence it is feasible to test a possible dependence of the extent
of galaxy dark matter halos on the density of the environment by binning the
data appropriately.  

\begin{figure}
  \begin{center}
    \epsfysize=4.5cm
    \epsffile{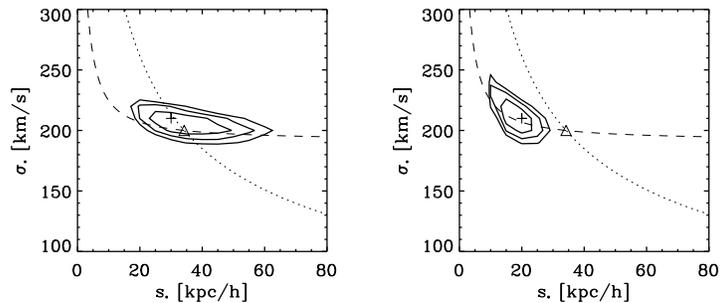}
  \end{center}
\caption[]{\small The results of the likelihood analysis for a cluster galaxy
input model with an extended dark matter halo. Note the change of scale on the
x-axis compared to the previous figures. The binning of the data into
information coming from images in the outer regions of the cluster ({\bf left})
and the cluster centre ({\bf right}) is the same as in Fig.\,\ref{small}.}
\label{large}
\end{figure}

Figure\,\ref{large} displays the results for the same realization of background
galaxies as above -- but here a large input value was used for the cutoff
radius of the cluster galaxies ($s_{\star}=34h^{-1}\,\rm{kpc}$). The left plot,
calculated from images in the outer region of the cluster, shows that in this
case the velocity dispersion can be reasonably well determined, whereas the
lensing effects are less sensitive to the radial extent of the mass
distribution. Nevertheless a robust lower limit of about $15h^{-1}\,\rm{kpc}$
can be set for the cutoff radius, and so this model can be distinguished with
high significance from the low-$s_{\star}$ model used above. 
Figure\,\ref{large} (right) reveals the problems of the method when it is
applied to the images located in the cluster centre. The input values cannot be
reproduced by the likelihood analysis in this case. The reason for this is the
ambiguity introduced by the mass correction procedure referred to in
Sect.\,\ref{method}. This problem does not show up for the input model with
small cutoff radius because then the mass in galaxies only amounts to a few
percent of the total mass, and the mass correction is not important. In the
outer regions of the cluster, the problem is less severe, because the
requirements for the accuracy of the cluster mass reconstruction are less
stringent in the weak lensing regime when the surface mass density is low and
so the method works there even when the galaxies are massive
(Fig.\,\ref{large} left). In the non-linear lensing regime of the cluster
centre, however, an accurate description of the cluster mass distribution is
essential to obtain reliable results.   

In order to solve the problem becoming apparent in Fig.\,\ref{large} (right)
one might envisage employing a maximum likelihood reconstruction of
the cluster mass distribution in the fashion of Bartelmann et al. (1996). In
such a method the presence of cluster galaxies could be taken into account
explicitly during the reconstruction process. For each set of parameters of
the galaxy mass model, one can then determine the best representation of the
underlying cluster mass distribution. Therefore, this approach would also be
more satisfactory in a full maximum likelihood sense. Finally, I would like to
remark that making the distinction between dark matter associated to galaxies
or belonging to a global cluster mass distribution becomes somewhat artificial
in the very centre of galaxy clusters when the physical distances between the
galaxies become very small, and clearly the giant cD-galaxies residing in the
centre of many clusters cannot be treated with the same formalism as ordinary
cluster galaxies.   

\subsection{Scaling Parameters}
In addition to $\sigma_{\star}$ and $s_{\star}$, a full description of the
model for the galaxy mass distribution also requires to specify the power
indices of the scaling relations (\ref{scal}). In the previous subsection the
same values ($\eta=4$, $\nu=0.5$) that had been used to generate the data were
taken for the likelihood analysis, but giving up that restriction and varying
the scaling parameters within reasonable ranges does not affect the general
conclusions drawn there.  

\begin{figure}
  \begin{center}
    \epsfysize=4.5cm
    \epsffile{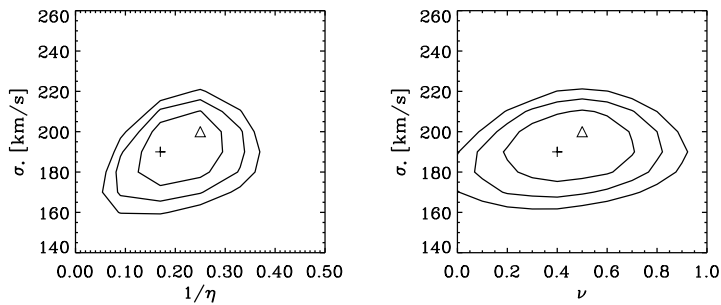}
  \end{center}
\caption[]{\small The dependence of the likelihood function on the scaling
indices $\eta$ and $\nu$. Again, the contours are $\Delta l=-1,-2,-3$, the
triangles denote the input values and the crosses mark the maximum of the
likelihood function.} 
\label{indices}
\end{figure}

Here the prospects for determining these scaling indices from the lensing
analysis are briefly mentioned. For generating the data the galaxy model
with the small cutoff radius has been adopted. Figure\,\ref{indices} depicts
contour plots for the logarithm of the likelihood as a function of $1/\eta$ and
$\sigma_{\star}$ (left) and $\nu$ and $\sigma_{\star}$ (right) including the
background images from the whole field of view. Each time the two remaining
parameters were fixed at the input values. The plots show that the constraints
on the scaling indices are not particularly tight. In order to improve them it
would be necessary to add the information from several galaxy clusters.

\section{Prospects}\label{prospects}
\subsection{Simulations}
There are several ways in which the simulations presented here could be
refined. A distinction should be made between spiral and elliptical cluster
galaxies because they require different normalizations for the velocity
dispersion. An obvious thing to do is to explicitly include a dependence of the
cutoff radius as a function of the (three-dimensional) density of the
environment. One can then develop strategies to quantify this dependence and to
assess the uncertainties introduced by projection effects. For this study it
was assumed that cluster galaxies and background galaxies can be unambiguously 
distinguished by means of some colour criterion. The importance of this
assumption can be tested by deliberately misinterpreting faint cluster galaxies
as background galaxies. First investigations in this direction indicate that
this is a minor problem, because the main part of the signal is contributed by
the more massive cluster galaxies.
\subsection{Observations}
Weak lensing is a challenging project from the observational point of view,
because it necessitates measuring accurate image ellipticities for a large
number of faint galaxies. To achieve the galaxy number density of
$40/\rm{arcmin}^2$ used in this simulations requires deep observations with a
magnitude limit of about $25$. The unique image quality of the
(refurbished) Hubble Space Telescope allows to determine image ellipticities
with a high accuracy, and in this respect it is the ideal instrument for weak
lensing purposes. Its drawback, on the other hand, is the rather small field of
view of its `wide field' camera, and so time consuming mosaics of several
images are required in order to completely cover a cluster which is located at
a reasonable redshift. However, it has been shown in recent years that
ground-based observations can be used for weak lensing studies as well,
provided that they were taken in good seeing and with telescopes and
instruments whose imaging properties are sufficiently well understood.
Several observations -- from space as well as from the ground -- which are
suitable for carrying out the kind of analysis described in this contribution
are already available, and clearly this will be a rewarding project for the
VLT-era. 

\bigskip
\noindent
{\bf Acknowledgments.} I thank Matthias Bartelmann for making the N-body
cluster simulation available, Peter Schneider and Stella Seitz for
discussions and for providing the reconstruction algorithm, and Matthias and
Peter for comments on the manuscript. This work was supported in part by the
Sonderforschungsbereich 375-95 der Deutschen Forschungsgemeinschaft. 

\end{document}